\documentclass{PoS}
\usepackage{graphics}
\usepackage{epsfig}
\usepackage{amssymb}
\usepackage{cite}
\usepackage{subfigure}
\usepackage{mcite}

\title{Experience with CMS pixel software commissioning}
\ShortTitle{Experience with CMS Pixel software commissioning}

\author{\speaker{V.~Chiochia}\thanks{On behalf of the CMS Pixel Group.}\\
        Physik Institut, Universit\"at Z\"urich, 8057 Z\"urich, Switzerland\\
        E-mail: \email{vincenzo.chiochia@cern.ch}}

\abstract{The CMS Pixel detector, consisting of three barrel layers and two endcap disks at each barrel end, was installed in the CMS experiment in summer 2008. After a preliminary commissioning phase with pulse injections the detector participated to data taking with cosmic ray trigger and 3.8T field. In this article, we describe the software for event reconstruction, as well as the status of the calibration tools. In addition, we report on the first running experience with CMS and present preliminary results on detector performance.}

\FullConference{17th International Workshop on Vertex detectors\\
		 July 28 - 1 August  2008\\
		 Ut\"o Island, Sweden}

\begin{document}

%%% INTRODUCTION %%%
\section{Introduction}

The CMS experiment, at the Large Hadron Collider (LHC) includes a silicon pixel detector to allow tracking in the region closest to the interaction point~\cite{:2008zzk}. The detector, installed in July 2008, will be a key component for re\-con\-struc\-ting interaction vertices and heavy quark decays in a particularly harsh environment, characterized by a high track multiplicity and heavy irradiation.

The detector consists of three barrel layers and two end-cap disks at each barrel end. The innermost barrel layer has a radius of 4.3 cm, while for the second and third layers the radius is 7.2 cm and 11 cm, respectively. The layers are composed of modular detector units. These modules consist of thin, segmented silicon sensors with highly integrated readout chips connected by the bump bonding technique. They are attached to cooling frames, the cooling tubes being an integral part of the mechanical structure. The barrel region is composed of 672 full modules and 96 half modules. The number or readout channels per module is 66'560 (full modules) or 33'280 (half modules). 

The endcap disks extending from about 6 to 15 cm in radius are placed at $z=\pm34.5$ cm and $z=\pm46.5$ cm. Disks are divided into half-disks including 12 "U"-shaped cooling channels disposed in a turbine-like geometry to enhance charge sharing. The endcap part includes 672 detector modules of five different sizes.

The minimal pixel cell size is dictated by the readout circuit area required for each pixel. In finding and localizing secondary decay vertices both transverse ($r\phi$) and longitudinal ($z$) coordinates are important in the barrel region. Therefore a nearly square pixel shape is preferred. Because charge is often shared among several pixels, the use of analogue signal readout enables position interpolation improving the spatial resolution. In the barrel the charge sharing in the $r\phi$-direction is large due to the 4 T magnetic field of the CMS solenoid. With a sensitive detector thickness of 285 $\mu$m the pixel size is 100 $\mu$m and 150 $\mu$m along the $r\phi$ and $z$ coordinates, respectively~\cite{Allkofer:2007ek}. 

One of the greatest challenges in the design of the pixel detector is the high radiation level on all components at very close distances to the colliding beams. At full LHC luminosity the innermost barrel layer will be exposed to a particle fluence of $3\times10^{14}$ n$_{\rm{eq}}/$cm$^2$/yr, the second and third layer to about $1.2\times10^{14}$ n$_{\rm{eq}}/$cm$^2$/yr and $0.6\times10^{14}$ n$_{\rm{eq}}/$cm$^2$/yr, respectively\footnote{All fluences are normalized to the non-ionizing energy loss (NIEL) of 1 MeV neutrons (n$_{\rm{eq}}/$cm$^2$).}. All components of the pixel system are specified to stay operational up to a particle fluence of at least $6\times10^{14}$ n$_{\rm{eq}}/$cm$^2$. 

The detector performances are expected to evolve with the exposure to irradiation and variation of the electric field across the sensor bulk. It is well known that sensor bias voltage will have to be increased with increasing irradiation to compensate the charge losses due to trapping. Thus, the event reconstruction software is designed to cope with a varying charge collection efficiency and precisely measure the hit position throughout the detector lifetime. The reconstruction techniques rely on periodical calibration procedures.

This paper describes the calibration and reconstruction software for the CMS Pixel detector and well as early results from data collected with cosmic ray trigger. The paper is structured as follows: The hit reconstruction techniques are discussed in Section~\ref{sec:hit_reconstruction}. A description of the offline calibration procedures is given in Section~\ref{sec:calibrations}. First results from data collected with cosmic ray trigger are presented in Section~\ref{sec:results}. The conclusions are given in Section~\ref{sec:conclusions}.

%%% RECONSTRUCTION %%%
\section{Hit reconstruction\label{sec:hit_reconstruction}}

The position of the pixel clusters is calculated in a two-step procedure using different reconstruction algorithms. During track seeding and pattern recongnition the hit position is calculated using the algorithm described in Section~\ref{sec:CPEGeneric}. This algorithm is fast but does not provide the ultimate precision, therefore, in the final track fit the hit position is recalculated using a more precise algorithm based on cluster shapes. The so-called {\it template} technique is described in Section \ref{sec:Templates}. This section shows also a comparison of the position resolution of both techniques based on the CMSSW simulation software.

%% CPE Generic
\subsection{First-pass hit reconstruction\label{sec:CPEGeneric}}

\begin{figure}[hbt]
  \begin{center}
    \scalebox{0.60}{\epsfig{file=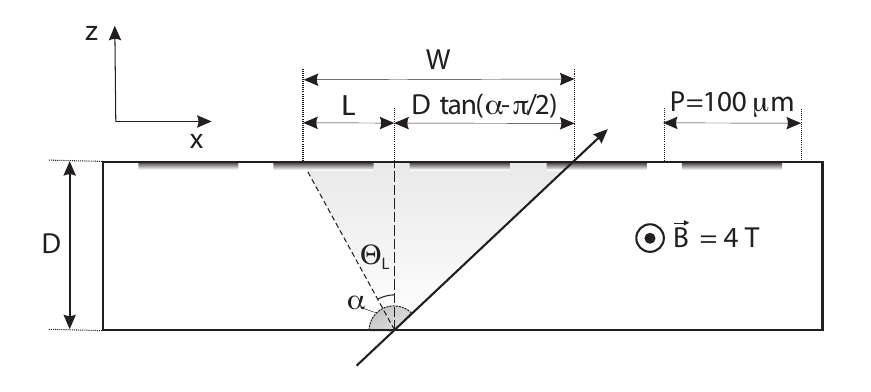,width=\linewidth}}
    \caption{Determination of the impact position in the transverse plane.}
    \label{fig:pos_resolution} 
  \end{center}
\end{figure}

Pixel clusters are formed by adjacent pixels with charge above the readout threshold. Both side and corner adjacent pixels are included in the cluster and a cut of 5'000 electron is applied on the total charge. The cluster is projected along the transverse ($x$) and longitudinal ($y$) direction by summing the charge collected in the pixels with the same coordinate. The Lorentz effect enhances the cluster lenght along the $x$ coordinate by $L=D\tan{(\Theta_L)}$ . For a cluster size of one the position is given by the center of the hit pixel corrected for the Lorentz shift. For larger clusters the hit position is determined by the expressions:
\begin{eqnarray}
x  &=& x_C + \frac{Q_{\rm last}-Q_{\rm first}}{2(Q_{\rm last}+Q_{\rm first})}|W_x-W^x_{\rm inner}|-\frac{1}{2}L,\label{eq:pos_reconstruction1}\\
y  &=& y_C + \frac{Q_{\rm last}-Q_{\rm first}}{2(Q_{\rm last}+Q_{first})}|W_y-W^y_{\rm inner}|,
\label{eq:pos_reconstruction2}
\end{eqnarray}
where $Q_{\rm first}$ ($Q_{\rm last}$) is the charge in electrons collected in the first (last) pixel above threshold, $(x_C,y_C)$ is the geo\-me\-tri\-cal center of the cluster and $W_{x,y}$ is the the total charge width given by:
\begin{eqnarray}
W_x &=& D \tan(\alpha-\frac{\pi}{2}) + L, \\
W_y &=& D \tan(\beta-\frac{\pi}{2}).
\label{eq:charge_width}
\end{eqnarray}
$W_{\rm inner}^{x,y}$ represents to the inner lenght of the cluster and is equal to (size-2)$\cdot$pitch.
Fig.~\ref{fig:pos_resolution} shows the definition of the track impact angle $\alpha$
with respect to the sensor plane along the $x$ direction. Similarly, $\beta$ corresponds to the impact angle along the $y$ coordinate.

%% Templates
\subsection{Clusters template hit reconstruction\label{sec:Templates}}

The template-based reconstruction algorithm is a procedure that translates pre-stored cluster projection shapes, also called ``templates'', across measured cluster projections to find the best fit and hence an estimate of the hit position in both transverse and longitudinal coordinates~\cite{Swartz:2007zz}.  The Pixelav simulation~\cite{Swartz:2003ch} is used to generate the templates which are stored as functions of the impact angles along with large quantities of auxiliary information in a template object. The simulation was originally written to interpret beam test data from several unirradiated and irradiated sensors.  It was extremely successful in this task, demonstrating that simple type inversion is unable to describe the measured charge collection profiles in irradiated sensors and yielding unambiguous observations of doubly-peaked electric fields in those same sensors \cite{Chiochia:2004qh}.

During the first phase of template generation,  the transverse ($x$) and longitudinal ($y$) projections for each simulated cluster are summed into respective 7-pixel and 21-pixel arrays.  By construction, the hit pixel is the center pixel in each projection.  The $x$ and $y$ coordinates of the hit are each binned in bins of width 0.125 pixel pitch.  The bins are chosen so that the middle bin is centered on the pixel center and the end bins are centered on the pixel boundaries.  This yields 9 bins spanning the central pixel where the end pixels differ by a full pixel pitch. The total charge, square of the charge, and number of entries are summed for each of the 9 (8 independent) bins.  The template consists of the average signal $S^{y/x}_{i,j}$ in each projected pixel $i$ and bin $j$.
\begin{figure*}[!htb]
\begin{center}
\resizebox{0.99\linewidth}{!}{\includegraphics{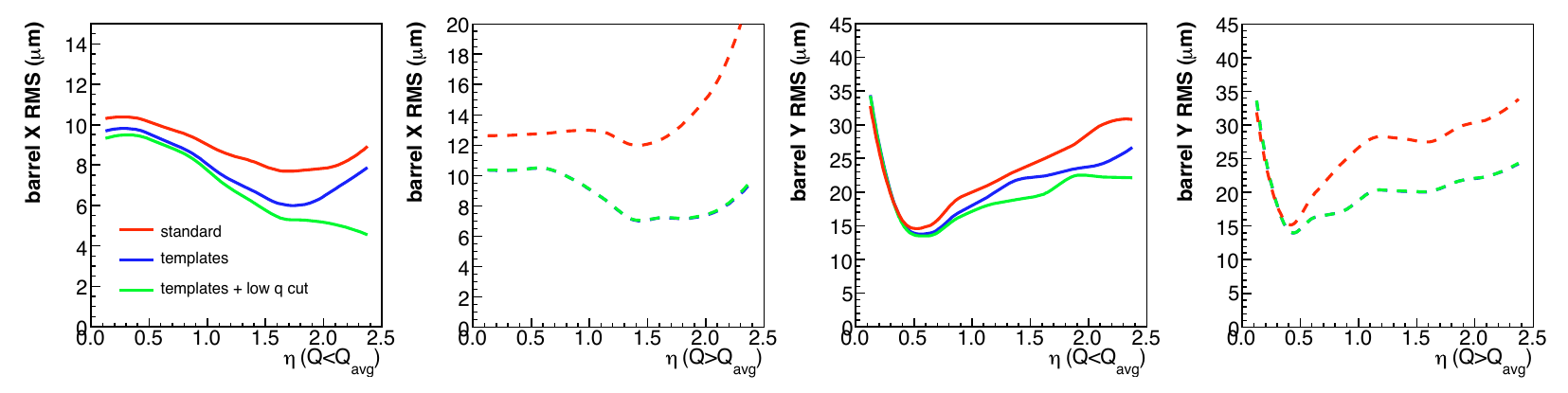}}
\caption{
The rms of the barrel transverse (local $x$ coordinate) residuals and longitudinal (local $y$ coordinate) of a CMSSW-generated sample of standard- and template reconstructed barrel clusters are plotted versus pseudorapidity for the cluster charge bands $1.5 > Q/Q_{avg} > 1$ and $1 > Q/Q_{avg}$. The red curves show the first-pass (standard) algorithm, the blue curves show the template algorithm and the green curves show the template algorithm after the removal of the low charge clusters. Note that the green and blue curves are coincident for the $1.5 > Q/Q_{avg} > 1$ band.
}
\label{template_res}
\end{center}
\end{figure*}

The second phase of the template generation uses the pre-stored results of the first phase to apply the actual template reconstruction algorithm to the same data samples used to generate the 9-bin templates.  The second pass generates information on biases, errors, corrections, and goodness-of-fit that are combined with the results of the first pass to build a 448~kB template summary file that represents a given set of operating conditions as simulated by Pixelav.

The basic goal of the procedure is to translate the expected cluster shape until it best matches the observed cluster shape.  This is accomplished by evaluating the following chisquare function for some or all of the template bins,
\begin{eqnarray}
\chi^2(j) &=& \sum_i \frac{(P^{y/x}_i - N_jS^{y/x}_{i,j})^2}{(\Delta P^{y/x}_i)^2}  \label{eq:chione} \\
            N_j &=& \sum_i \frac{P^{y/x}_i}{(\Delta P^{y/x}_i)^2}\bigg{/} \sum_i \frac{S^{y/x}_{i,j}}{(\Delta P^{y/x}_i)^2} \nonumber
\end{eqnarray}
where $P^{y/x}_i$ are 1-d projections of the {\it measured} cluster and $\Delta P^{y/x}_i$ the respective rms. A more complete description of the template reconstruction technique can be found in~\cite{Swartz:2007zz}.

The template algorithm was tested using samples of events with six 20 GeV muons generated by the CMSSW simulation software. A comparison of the template and first-pass algorithms is shown in Fig.~\ref{template_res}. The rms resolutions in the two charge bands are shown in the standard algorithm (red lines), the template algorithm (blue lines), and the template algorithm after the low charge clusters have been removed with a simple charge cut (green lines). Note that the template algorithm still outperforms the standard one with all effects present and improves further when the low charge clusters are removed.

%%% CALIBRATIONS %%%
\section{Offline calibrations\label{sec:calibrations}}

A precise hit position determination relies on the correct calibration of the detector response. The CMS pixel hit reconstruction requires two sets of calibration constants: the front-end gain constants and the amplitude of the Lorentz angle. The former is needed to convert the pixel analog charge from ADC counts into electrons. The latter is used to calculate Eqs.~\ref{eq:pos_reconstruction1}-\ref{eq:pos_reconstruction2}. The two calibration procedures are described in the following sections.

%%% Gain Calibration %%%
\subsection{Calibration of the front-end gain response\label{sec:gain_calibration}}
\begin{figure}[hbt]
  \begin{center}
     \mbox{
      \subfigure[]{\scalebox{0.325}{
	  \epsfig{file=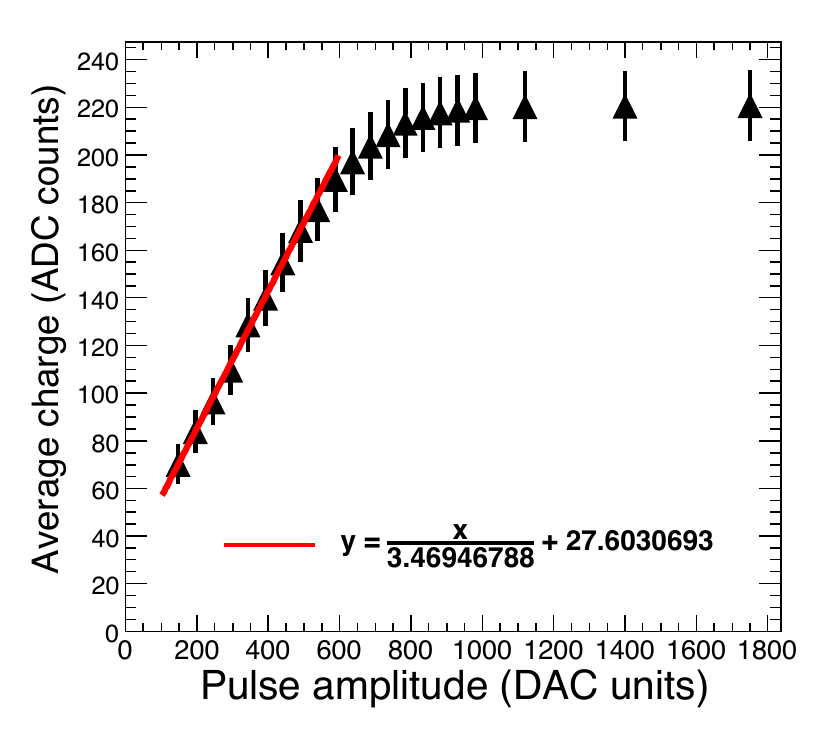,width=\linewidth}
	  \label{fig:gaincurve} 
      }}
    }
    \mbox{
      \subfigure[]{\scalebox{0.28}{
	  \epsfig{file=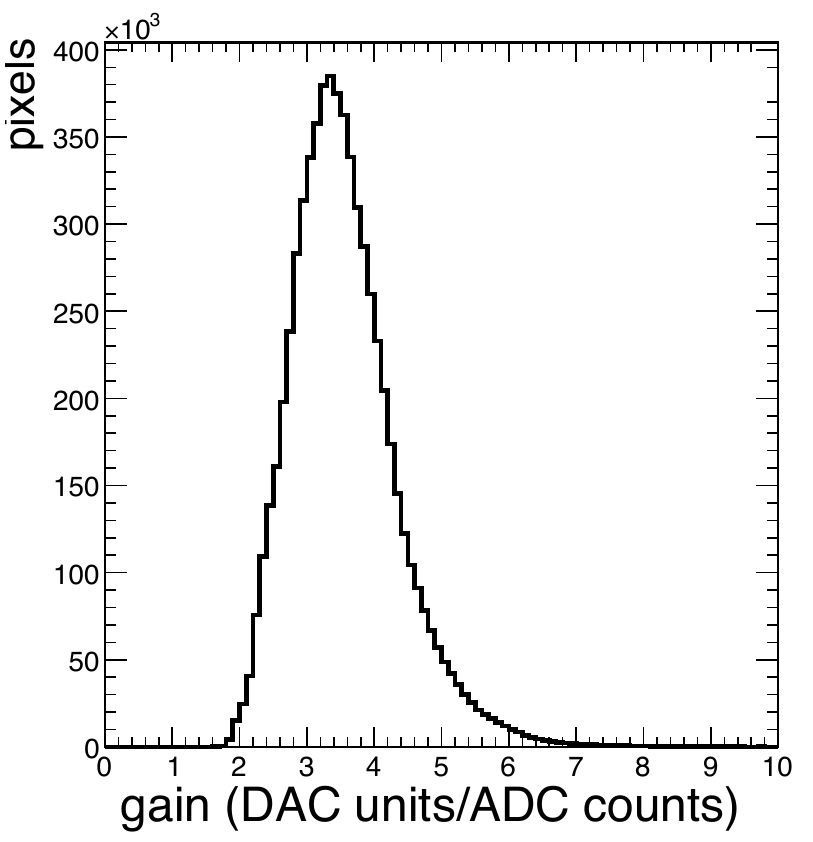,width=\linewidth}
	  \label{fig:gainvalue} 
      }}
    }
    \mbox{
      \subfigure[]{\scalebox{0.285}{
	  \epsfig{file=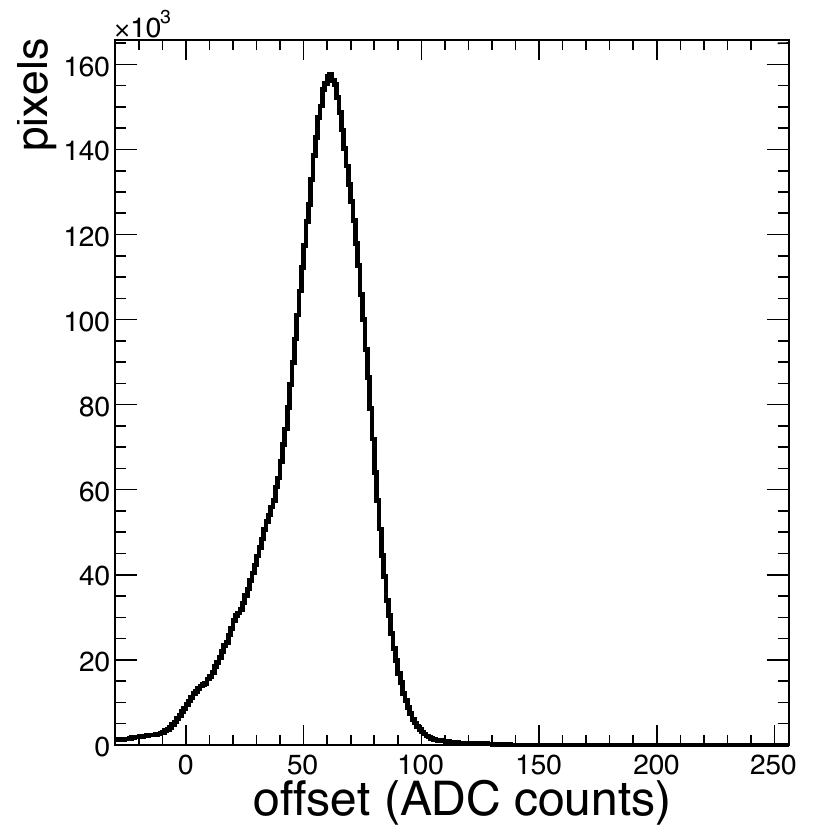,width=\linewidth}
	  \label{fig:pedestalvalue} 
      }}
    }
    \caption{(a) Gain response function for a single pixel. The solid red line shows a straight-line fit to the linear region. Gain (b) and offset (c) distribution extracted from straight-line fits to the gain response curves of all pixels.}
  \end{center}
\end{figure}

The measurement of the gain response is performed for each pixel readout chip (ROC) cell in two steps. Firstly, the gain response is measured by injecting a pulse of variable amplitude in each cell of the readout chip (ROC). The injection is performed several times for each pulse amplitude and the average charge in ADC counts is computed for each point. The gain response function of a single cell is shown in Fig.~\ref{fig:gaincurve} as function of the injected pulse amplitude in DAC units. The front-end response is linear up to about 600 DAC units and then saturates. The saturation is beneficial for suppressing large Landau fluctuations and charge deposits due to secondary electrons.
%\begin{figure}[hbt]
%  \begin{center}
%    \scalebox{0.50}{\epsfig{file=GainCurve_Good.pdf,width=\linewidth}}
%    \caption{Gain response function for a single pixel. The solid red line shows a %straight-line fit to the linear region.}
%    \label{fig:gaincurve} 
%  \end{center}
%\end{figure}

The second step consists in performing a straight line fit to the linear part of the gain curve and converting DAC units into electrons. Offset and gain (i.e. the inverse of the fit slope) from the fit are stored in a ROOT file for each pixel cell, together with the fit $\chi^2$, its probability and the number of d.o.f. The fit procedure is successful for 99.8\% of the pixels. About 0.2\% of the pixels show a saturated gain response over the full amplitude range and the fit fails. The $\chi^2$ probability is larger than 0.1\% for about 97\% of the pixels. A visual check of the pixels with $P(\chi^2)<0.1\%$ shows deviations from linear trend at low input amplitudes. The average offset and gain from the rest of the pixels is assigned to the cells where the fit does not converge properly or a $P(\chi^2)<0.1\%$ is returned. 
The slope and offset distribution for all pixel cells are shown in Fig.~\ref{fig:gainvalue} and \ref{fig:pedestalvalue}, respectively. The relative spread (rms divided by mean value) is 37\% for the offsets and 22\% for the slopes.

The gain calibration constants are stored in the CMS offline database and used during event reconstruction. To contain the memory needs of the event reconstruction process the gains are averaged over ROC columns (80 channels) while pixel-granularity is used for the offsets. The final payload size is 67~MB. At the High Level Trigger (HLT) both offsets and gains are averaged over ROC columns, for a total payload size of about 2~MB. 
%Studies based on detector simulation have shown that column granularity of the gain calibration constants deteriorates the position resolution in the transverse plane by XX\%, however, it considerably reduces the activation time of the HLT at run startup. 
While the reduced payload is acceptable for triggering purposes the best position resolution is achieved during offline event reconstruction due to the high granularity. 

The measured pixel charge in ADC counts is converted into electrons by applying the calibration constants discussed above and converting DAC units into electrons. This last step is performed by applying the DAC-to-electrons conversion constants measured during module production with a x-ray source~\cite{TruebThesis}.

%%% Lorentz Angle %%%
\subsection{Lorentz angle calibration}
Due to the dependence of the charge carrier mobility on temperature and detector exposure to radiation a periodic measurement in-situ of the Lorentz angle is highly desirable. During operations with colliding beams the angle can be measured with the {\it grazing angle} method as described in~\cite{Wilke:2008zz}. The method relies on the measurement of the average cluster deflection with respect to the direction of the incoming track. It was first applied on data collected with a pion beam and prototype sensors at CERN~\cite{Allkofer:2007ek} and later optimized for tracks from proton collisions using CMSSW detector simulations. As reported in~\cite{Wilke:2008zz} the Lorentz angle can be measured with a 2\% accuracy with 1'000 tracks. The high track rate during collisions allows to measure the value in different regions of the detector, e.g. as function of the detector rapidity. A different value of the Lorentz angle can be expected in various parts of the detector due to their different exposure to irradiation.

Data collected with cosmic ray trigger lead to tracks crossing the detector with a wide range of impact angles and a different measurement technique can be applied. The technique relies on the measurement of the cluster size as function of the impact angle in the transverse plane. The cluster width in the transverse plane is minimal when the impact angle is parallel to the direction of the Lorentz deflection and is larger for all other angles. Thus, the Lorentz angle amplitude can be determined by measuring the minimum of the cluster size distribution. 

%%% FIRST RESULTS %%%
\section{First results from cosmic ray data\label{sec:results}}
The pixel detector was included in the CMS data taking period with cosmic ray trigger in autumn 2008. About 94\% of the end-cap wheels and 99\% of the barrel part were included in the run. The missing section of the end-cap wheel was mainly due to a short in a power distribution card developed after installation which will be repaired during the next winter shutdown. 

Data were recorded both with and without 3.8~T magnetic field. Among 8.5 million tracks detected in the CMS silicon strip tracker about 77'000 tracks traversing the pixel detector volume were recorded. Noisy pixel cells where detected with the online data quality monitoring software~\cite{Dutta:2006cn} by applying a cut on the pixel frequency. In total, only 263 and 17 pixel cells were masked during data taking in the barrel and end-cap regions, respectively.
Tracks in the pixel and strip tracker were promptly reconstructed with three different cosmic track finding algorithms currently available in the CMSSW reconstruction software~\cite{Adam:2008zz}. The pixel hit reconstruction algorithms are the ones described in Section~\ref{sec:hit_reconstruction}. 

\begin{figure}[hbt]
  \begin{center}
    \mbox{
      \subfigure[]{\scalebox{0.41}{
	  \epsfig{file=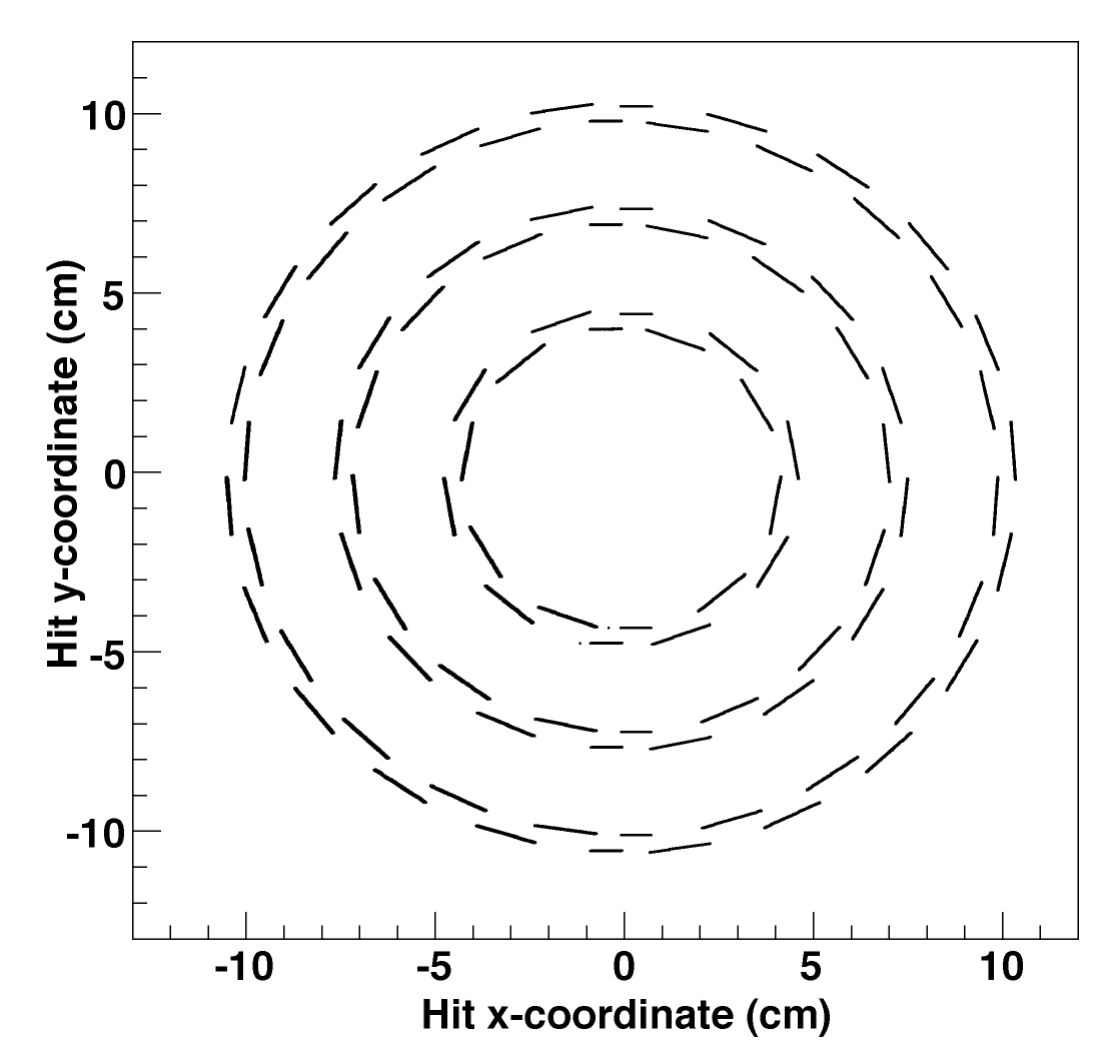,width=\linewidth}
	  \label{fig:hitmap} 
      }}
    }
    \mbox{
      \subfigure[]{\scalebox{0.44}{
	  \epsfig{file=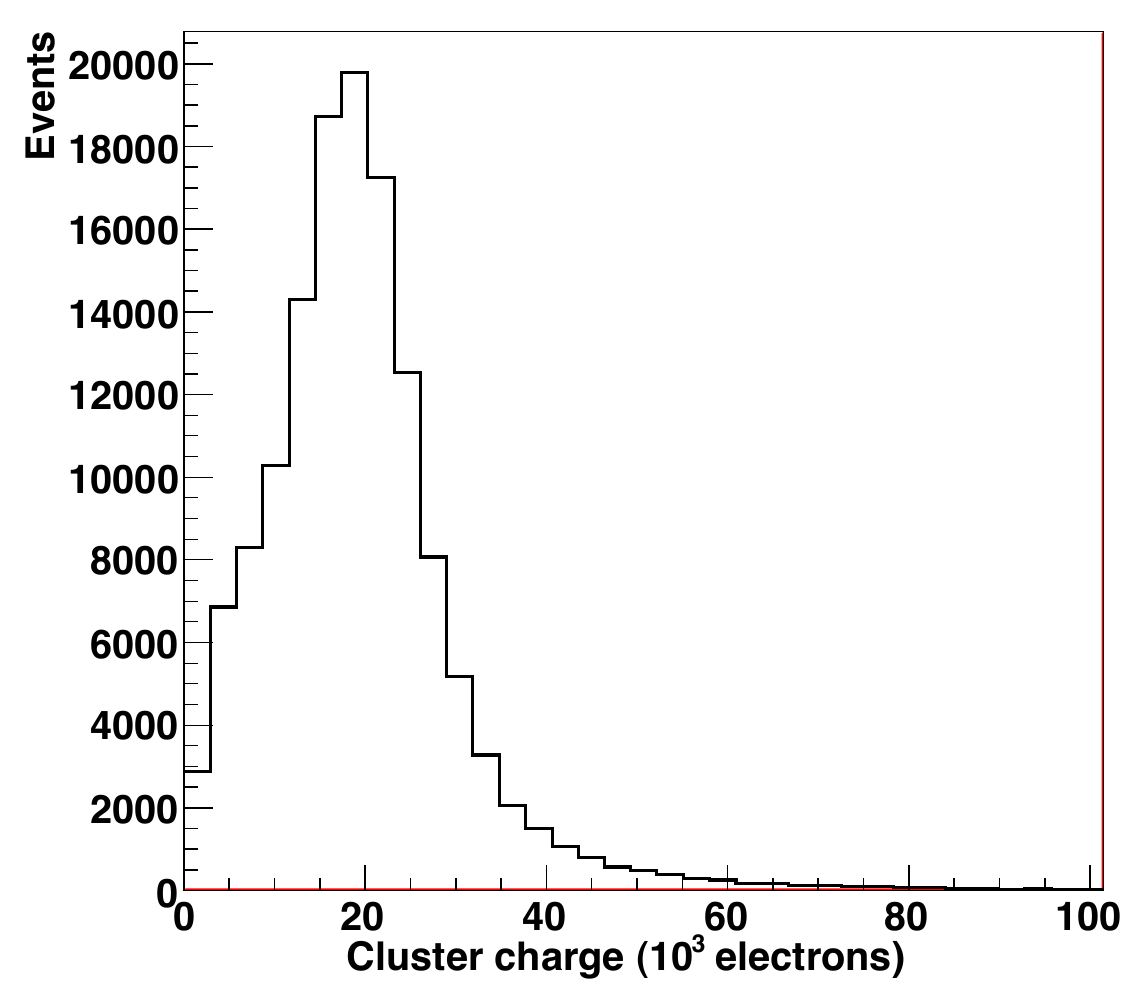,width=\linewidth}
	  \label{fig:BPIXcharge} 
      }}
    }
    \caption{(a) Barrel pixel hit distribution in the CMS transverse plane from cosmic ray data. (b) Cluster charge distribution in the barrel pixel detector corrected for the impact angle.}
  \end{center}
\end{figure}

The distribution of barrel hits in the transverse plane is shown in Fig.~\ref{fig:hitmap}. The hit position was found to be correct in all detector parts except the barrel half-modules of the negative $z$ region for which the local coordinates were inverted. A bug fix was deployed for the reprocessing of the data sample. The calibrated charge distribution corrected for the track impact angle is shown in Fig.~\ref{fig:BPIXcharge}. The distribution peak is in line with the charge deposit expected for a minimum ionizing particle traversing 285 $\mu$m of silicon. The measured mean charge and rms are $19.4\times10^3$ and $10.8\times10^3$ electrons, respectively. No pixel inter-calibration was applied during prompt event reconstruction and a unique set of offset and gain constants were used for all pixels. The result of the gain calibration scans are currently under study and will be applied in future data reprocessing.

The time alignment between the pixel detector and the trigger signal from the muon system was optimized during the run. The track detection efficiency in the pixel detector was measured with different time delays with respect to the first level trigger signal. The efficiency was defined as the ratio $N_{\rm detected}/N_{\rm expected}$ where $N_{\rm expected}$ is the number of tracks reconstructed in the strip tracker with transverse impact parameter and longitudinal parameter below 10~cm and 25~cm, respectively, and at least one hit in the bottom part of the micro-strip Tracker Inner Barrel. The numerator $N_{\rm detected}$ was defined by the number of tracks including at least one hit in pixel barrel region. At the optimal delay the measured efficiency is about 70\%. The fraction of undetected cosmic tracks can be explained by the longer shaping time set in the micro-strip front-end chips and by the jitter of the trigger signal.

%%% CONCLUSIONS %%%
\section{Conclusions\label{sec:conclusions}}

This paper has described the main aspects of the calibration and reconstruction software of the CMS pixel detector. Two reconstruction techniques were implemented for the measurement of the cluster position. A fast algorithm is used during track seeding and pattern recognition while a more precise and CPU intensive technique is applied during the final track fit to determine the track parameters. The latter, based on pre-computed average cluster shapes improves the hit position and can cope with charge trapping effects due to irradiation. The main offline calibration procedures consist in the measurement of the gain response and of the Lorentz angle amplitude. 

The reconstruction and calibration workflow was successfully tested during the CMS commissioning phase with cosmic muon trigger. About 77'000 tracks and 200'000 hits were detected in the pixel detector. The very first results presented in this paper show excellent performances of the detector in terms of signal and noise. The analysis of the full data-set is expected to be completed during the winter shutdown.

%%%%%%%%%%%%%%%%%%%%%%%%%%%%%%%%%% REFERENCES %%%%%%%%%%%%%%%%%%%%%%%%%%%%%%%%%

\end{document}